\documentclass{iopart}
\expandafter\let\csname equation*\endcsname\relax
\expandafter\let\csname endequation*\endcsname\relax
\usepackage{amsmath}
\usepackage{graphicx}
\usepackage{bm}
\usepackage[british]{babel}
\usepackage{braket}
\usepackage{siunitx}
\usepackage{hyperref}
\usepackage{tikz}
\usepackage[hyperref=true,
url=false,
maxbibnames=99,
isbn=false,
doi=false,
eprint=false,
backref=false,
backend=biber,
natbib=false,
giveninits=true,
style=numeric-comp,
sorting=none
]{biblatex}
\AtEveryBibitem{\clearfield{month}}
\AtEveryBibitem{\clearfield{day}}
\AtEveryBibitem{\clearfield{pages}}
\AtEveryBibitem{\clearfield{pagetotal}}
\AtEveryBibitem{\clearfield{issn}}
\AtEveryBibitem{\clearfield{address}}
\usepackage{xcolor}
\usepackage{csquotes}
\usepackage{changes}
\usepackage{amssymb}
\usepackage{iopams}

\DeclareSourcemap{
  \maps[datatype=bibtex]{
    \map[overwrite]{
      \step[fieldsource=journal, fieldtarget=journaltitle]
    }
  }
}

\begin{document}
\title{Modelling spectra of hot alkali vapour in the saturation regime}
\author{Daniel R Häupl\textsuperscript{1,2}, Clare R Higgins\textsuperscript{3}, Danielle Pizzey\textsuperscript{3}, Jack D Briscoe\textsuperscript{3}, Steven A Wrathmall\textsuperscript{3}, Ifan G Hughes\textsuperscript{3}, Robert Löw\textsuperscript{4}, Nicolas Y Joly\textsuperscript{1,2}}
\address{\textsuperscript{1}University of Erlangen-Nürnberg, Staudtstraße 7/B2, 91058 Erlangen, Germany}
\address{\textsuperscript{2}Max Planck Institute for the Science of Light, Staudtstraße 2, 91058 Erlangen, Germany}
\address{\textsuperscript{3}Department of Physics, Durham University, South Road, Durham DH1 3LE, United Kingdom}
\address{\textsuperscript{4}5\textsuperscript{th} Physical Institute, University of Stuttgart, Pfaffenwaldring 57, 70569 Stuttgart, Germany}

\date{\today}

\begin{abstract}
Laser spectroscopy of hot atomic vapours has been studied extensively.  Theoretical models that predict the absolute value of the electric susceptibility are crucial for optimising the design of photonic devices that use hot vapours, and for extracting parameters, such as external fields, when these devices are used as sensors.  To date, most of the models developed have been restricted to the weak-probe regime. However, fulfilling the weak-probe power constraint may not always be easy, desired or necessary. Here we present a model for simulating the spectra of alkali-metal vapours for a variety of experimental parameters, most distinctly at intensities beyond weak laser fields. The model incorporates optical pumping effects and transit-time broadening. We test the performance of the model by performing spectroscopy of $^{87}$Rb in a magnetic field of \SI{0.6}{\tesla}, where isolated atomic resonances can be addressed. We find very good agreement between the model and data for three different beam diameters and a variation of intensity of over five orders of magnitude. The non-overlapping absorption lines allow us to differentiate the saturation behaviour of open and closed transitions.  While our model was only experimentally verified for the D2 line of rubidium, the software is also capable of simulating spectra of rubidium, sodium, potassium and caesium over both D lines. 
\end{abstract}

\submitto{\NJP}
\maketitle
\section{Introduction}
\label{sec:intro}
Spectroscopy is an essential tool in fundamental research to unveil the electronic structure of atoms and molecules, through their interaction with radiation. Its powerful capability extends to many applications such as time and frequency referencing~\cite{piester2011ars, riley2019iuh} and quantum sensing of electro-magnetic fields~\cite{degen2017rmp}. Operating below saturation is crucial for obtaining a linear response and minimising perturbations caused by the radiation itself. At larger intensities, nonlinear effects arise. This regime can be utilised, for instance, to study a single atom~\cite{schuster2008np}.

Laser spectroscopy of hot atomic vapours has been studied extensively, with topics of interest including constructing a magnetometer~\cite{fabricant2023build}; electromagnetically induced transparency~\cite{finkelstein2023practical, zhang2024interplay}; terahertz imaging~\cite{downes2023practical}; nonlinear and quantum optics~\cite{glorieux2023hot}; dipolar interactions and cooperative effects in confined geometries~\cite{alaeian2024manipulating}; and building an optical filter~\cite{uhland2023build}; to name but a few.

Vapours of alkali metals are ideally suited media for many applications as they combine: (i) a large resonant optical depth~\cite{pizzey2022laser}; (ii) simple atomic structure with well-understood interactions with external fields~\cite{downes2023simple}; and (iii) resonance transitions in the near infrared for which there exist numerous convenient laser sources~\cite{mausezahl2024tutorial}.  
The simplicity of the theoretical framework and experimental apparatus results in alkali-metal vapours being the platform of choice for numerous experiments.
In the particular case of alkali-metal atoms probed by weak intensity laser fields, well-established theoretical models have been implemented in open-source software such as $ElecSus$~\cite{zentile2015cpc, keaveney2018cpc}. Its versatile design allows precise fitting to experimental atomic spectra to extract a large variety of experimental parameters, e.g. temperature, buffer gas broadening, isotope ratio or magnetic field strength, to list a few. However, fulfilling the weak-probe power constraint~\cite{siddons2008jpbamop} may not always be easy, desired or necessary. For certain experimental setups, involving e.g. waveguides~\cite{jones2014saturation, hickman2014saturated, song2019absorption, haupl2022njp}, cavities~\cite{schuster2008np}, nano-photonics~\cite{skljarow2020oeo} or when exploiting the nonlinear regime~\cite{mulchan2000josabj}, it may even be very challenging to operate below the saturation intensity. 
Above saturation, a variety of other effects need to be considered. Besides the obvious nonlinear response and power broadening, additional contributions from transit time broadening and optical pumping arise. The effects of transients and short excitation pulses are also a concern but are not included here.  
In this work, we present a model for alkali-metal vapour spectroscopy beyond weak intensity laser fields. We take into account the effects listed above, including all accessible hyperfine states. The model predicts the expected spectrum of an alkali-metal vapour at high intensities for different experimental parameters. 
To verify its validity, we performed D2 line spectroscopy of a $^{87}$Rb vapour subject to an external magnetic field, investigating a broad parameter space.
The laser intensity covered 5 orders of magnitude, and different laser beam diameters were used to explore the effect of transit time broadening. 
This variety of parameters demonstrates the versatility of the model.

The rest of this paper is organised as follows: Section~\ref{sec:theory} outlines the theoretical model;  we describe the atomic basis used for the calculation, and the relevance of being in the hyperfine Paschen-Back regime. The Linblad master equation is used to evaluate the atomic response and incorporate excited state decay. In Section~\ref{sec:experiment} we provide details of the apparatus used to conduct the investigation. In Section~\ref{sec:results} we present and analyse the experimental results, and compare with our model. Finally, we conclude and present an outlook in Section~\ref{sec:outlook}.

\section{Theory}
\label{sec:theory}
The complex electric susceptibility, $\chi(\omega)$, encapsulates the response of a dielectric medium when subject to monochromatic radiation at angular frequency $\omega$.  The susceptibility describes the bulk response of the medium---the real part characterising  dispersive effects, and the imaginary component the absorption of the light.  
The macroscopic response can be related to the microscopic response  of individual atoms, averaged over the distribution of atoms in the medium~\cite{foot2005,  loudon2000}. 
We shall employ the formalism of the master equation to incorporate spontaneous emission~\cite{steck2007}.
Our model takes into account the transit time broadening arising from the finite time atoms spend traversing the laser beam. 
Since the light is attenuated as it propagates through the medium, our model also incorporates propagation effects. In the next sub-section we provide details of the atomic Hamiltonian, and in what basis the calculations are performed. 

\subsection{The atomic Hamiltonian.}
\label{subsec:hamiltonian}
We shall consider the D2 and D1 transitions $n$S$_{1/2}$~$\rightarrow$~$n$P$_{3/2, 1/2}$ in alkali metals, with $n$ being the principal quantum number of the valence electron.  
The $n$P$_{3/2}$ and $n$P$_{1/2}$ terms differ in energy due to the coupling between the electron's orbital angular momentum \textbf{L} and  spin angular momentum \textbf{S}.  Further splitting into the hyperfine structure occurs due to coupling between the electron's total angular momentum \textbf{J} and the nuclear angular momentum \textbf{I}~\cite{foot2005}.

The Zeeman interaction of alkali-metal atoms subject to an external magnetic field is well known~\cite{Woodgate1980, bransden}. 
In brief, for weak fields the total angular momentum  $\mathbf{F}$, and its projection $m_F$, are good quantum numbers; for stronger fields the interaction with the external field exceeds the internal coupling of the atom, and therefore $m_I$ and $m_J$ are the good quantum numbers (a qualitative discussion of the field needed is provided in Section~\ref{subsec:HPB}). 
The evolution of the good quantum numbers differs between the excited state and the ground state, as the hyperfine interaction is weaker for the excited state.
Thus, there are values of the magnetic field at which the good quantum numbers are different in the ground and excited states, or there are no good quantum numbers for one state. 
The values of the magnetic field at which these different regimes occur depend on the hyperfine interaction strength, and therefore the alkali-metal atom, and isotopes of the same atom.
Consequently, it becomes favourable to use numerical techniques for the diagonalisation of the Hamiltonian to obtain the eigenenergies (to calculate transition energies) and eigenstate decompositions (to calculate transition line-strengths).
This is the approach used in $ElecSus$~\cite{zentile2015cpc, keaveney2018cpc}.
The Hamiltonian matrix incorporates both the hyperfine structure and Zeeman interaction using the uncoupled basis $\vert m_L, m_S, m_I\rangle$, where $m_L$, $m_S$ and $m_I$ are respectively the projection of the electron's orbital angular momentum, electron spin, and nuclear spin. 

The transition frequencies are obtained from the matrix diagonalisation. The relative transition strengths are calculated by evaluating  the electric dipole matrix elements squared; the absolute values are calculated using atomic properties and fundamental constants, as detailed in Section~1.7 of~\cite{pizzey2022laser}.  The selection rules are particularly simple in the uncoupled basis: neither the electron spin nor the nuclear spin projections change 
during an electric dipole transition, therefore $\Delta m_S \equiv m_{S'}-m_S = 0$, $\Delta m_I \equiv m_{I'}-m_I = 0$, where the primed notation indicates the excited state. 
Upon optical excitation with a dipole transition, the orbital angular momentum does change, and there is a simple selection rule for its projection:  $\Delta L \equiv L'-L = \pm1$, and $\Delta m_L \equiv m_{L'}-m_L = \pm1,\,0$, corresponding to $\sigma^{\pm}$, and $\pi$ transitions, respectively.
We define the quantisation axis $z$ along $B$.  Electric field components  in the $xy$-plane drive optical transitions if and only if $\Delta m_{L} = \pm 1$. The $\sigma^+$ ($\sigma^-$) transition is  driven by left (right)-hand circular polarised beams propagating along the quantisation axis~\cite{f2f}. 
A component of the electric field oscillating along $z$ induces the $\Delta m_{L} = 0$, or $\pi$-transition, but will not be studied in this investigation.

\subsection{The  hyperfine Paschen-Back regime.}
\label{subsec:HPB}
The hyperfine Paschen-Back (HPB) regime is attained when the Zeeman shift exceeds the ground state hyperfine interaction.  Numerous experimental studies have been performed in the HPB regime with Rb~\cite{tremblay1990absorption, Weller2012a, sargsyan2012hyperfine, Zentile2014a, sargsyan2014hyperfine, Voigt5, briscoe2023voigt}. Recent highlights show the breadth of the utility of this regime, covering the fundamental: precision measurement of the excited state Landé g-factor~\cite{staerkind2023precision}; electromagnetically induced transparency and optical pumping in the HPB regime~\cite{mottola2023electromagnetically}; through to the technological: quantum memories~\cite{mottola2023optical}; to the applied: high-field optical  magnetometry for magnetic resonance imaging~\cite{staerkind2024high}. To  estimate the field necessary to gain access to the HPB regime we equate the magnetic dipole constant for the ground state, $A_{\rm HFS}$, to the Zeeman shift of a ground stretched state: $B_{\rm HPB}=A_{\rm HFS}/\mu_{\rm B}$, where  $\mu_{\rm B}$ is the Bohr magneton. 
For $^{87}$Rb this is evaluated to be \SI{0.24}{\tesla}~\cite{pizzey2022laser}. 
When performing spectroscopy of Rb  in the HPB regime the separation of the optical transitions  (the Zeeman shift)   exceeds the Doppler width.  Consequently, the spectrum is comprised of  isolated atomic resonances; this greatly simplifies the interpretation of the atom-light interaction as complications arising from overlapping resonances are eliminated.  This has led to  textbook demonstration of: ladder-EIT~\cite{whiting2017single}, V-EIT~\cite{higgins2021electromagnetically}, electromagnetically induced absorption~\cite{WhitingEIA}, and diamond four-wave mixing~\cite{Whiting:FWM}. Importantly for this study, investigating the role of saturation and optical pumping with open and closed atomic transitions will be significantly easier with isolated resonances, when open and closed transitions can be probed independently.

\subsection{Lindblad master equation}
The density operator $\rho=\sum_{i}p_i\ket{i}\!\bra{i}$ describes the mixed quantum state of an ensemble of atoms, with $p_i$ the probability to end up in the pure $\ket{i}$ state after a measurement.
As we will see later, each state $\ket{i}$ corresponds to a vector of the system's eigenbasis.
The diagonal element of the corresponding density matrix $\rho_{ii}=\braket{i|\rho|i}$ corresponds to the population of $\ket{i}$ and the off-diagonal element $\rho_{ij}=\braket{j|\rho|i}$ to the coherence between the states $\ket{i}$ and $\ket{j}$~\cite{loudon2000}.

The theoretical basis of our semi-classical model are the well-known optical Bloch equations \cite{sokolov2002, loudon2000}.
There, the atom is considered a quantum-mechanical object, while the electromagnetic wave is approximated by a classical plane wave.
In this work, the optical Bloch equations are formalised by the master equation in the Lindblad form~\cite{manzano2020aa, steck2007}, describing open quantum systems, i.e. systems interacting with their environment:
\begin{equation}
    i\hbar\frac{\partial \rho}{\partial t}=\left[H, \rho\right]+i\hbar L(\rho).
    \label{eq:master_equation}
\end{equation}
The equation describes the temporal evolution of the density matrix in the presence of decoherence processes.
Here, $L(\rho)$ is the Lindblad operator and incorporates these decoherence processes, to represent quantum jumps by statistical decay or dephasing rates.
For a decay process it has the form~\cite{steck2007}
\begin{equation}
\begin{aligned}
    L(\rho) &= \sum_{i,j} (\Gamma_{ji}\rho_{jj}-\Gamma_{ij}\rho_{ii})\ket{i}\bra{i}\\
    &- \frac{1}{2}\sum_{i\neq j}\left(\sum_k \Gamma_{ik}+\Gamma_{jk} \right)\rho_{ij} \ket{i} \bra{j},
    \label{eq:lindblad_operator}
\end{aligned}
\end{equation}
with $\Gamma_{ij}$ the decay rate between state $\ket{i}$ and $\ket{j}$. 
$L_D(\rho)$ acts on both the coherences and populations.
An analogous form can be used to describe sole decoherence processes, e.g. for non-quenching collisions with a buffer gas~\cite{steck2007}.

The exact rate of $\Gamma_{ij}$ is given by the product of the squared dipole matrix element DME and the natural decay rate $\Gamma_{\mathrm{D}1/2}$ of the respective D line (here the D2 line)~\cite{steck2007} by
\begin{equation}
    \Gamma_{ij} = \left[ \mathrm{DME}^2_{\sigma^+} + \mathrm{DME}^2_{\sigma^-} + \mathrm{DME}^2_{\pi} \right] \cdot \Gamma_{\mathrm{D}1/2}.
\end{equation}
Here $\mathrm{DME}_{\sigma^+}$, $\mathrm{DME}_{\sigma^-}$, and $\mathrm{DME}_{\pi}$ correspond to the dipole matrix elements of the $\sigma^+$, $\sigma^-$ and $\pi$-transition. 
As described in Section~\ref{subsec:hamiltonian}, we separately derive the atomic Hamiltonian including the Zeeman interaction of ground and excited state in the completely uncoupled basis $\ket{m_L,m_S,m_I}$.
We then calculate the respective eigenbasis via diagonalisation.
This gives us a set of vectors, which allows the representation of the eigenbasis in terms of the uncoupled basis.
The dipole matrix elements are then given by the scalar product of the respective ground and excited state vectors.
In case of zero magnetic field the eigenbasis coincides with the completely coupled basis $\ket{F,m_F}$~\cite{foot2005}.
Transit time effects are factored in by an effective decay rate (see~\autoref{ssec:transit_time_broadening}).

For continuous wave lasers and due to the decoherence processes, our system is assumed to reach equilibrium.
Therefore the steady-state solution $\partial\rho/\partial t=0$ is sufficient.
With the steady-state solution of the density matrix, one obtains the absorption of the system via the electric susceptibility $\chi$.
Using the electric polarisation
\begin{equation}
    \mathbf{P} = \epsilon_0\chi \mathbf{E},
\end{equation}
we can link $\chi$ with the mean value of the atomic electric dipole $\braket{p_A}$~\cite{grynberg2010} via
\begin{equation}
    \mathbf{P} = n \braket{p_A} = n \sum_{ij}\rho_{ij}d_{ij}.
\end{equation}
Here $d_{ij}=\braket{j|e\mathbf{r}\cdot\hat{\epsilon}|i}$ is the dipole matrix element between state $\ket{i}$ and $\ket{j}$ of one atom, $n=N/V$ the atomic density, $\mathbf{r}$ is the position operator and $\hat{\epsilon}=\mathbf{E}/|\mathbf{E}|$ is the normalised polarisation vector of the external electromagnetic field $\mathbf{E}$.
This yields the relation
\begin{equation}
    \chi = \frac{n}{\epsilon_0 |\mathbf{E}|} \sum_{ij}\rho_{ij}d_{ji}
\end{equation}
between the electric susceptibility $\chi$ and the entries of the density matrix $\rho$.
The latter is obtained from numerically solving \autoref{eq:master_equation} in the steady-state.
\subsection{Transit time broadening}
\label{ssec:transit_time_broadening}
In thermal vapours, atoms have velocities following the Maxwell-Boltzmann distribution.
Therefore, an individual atom will only interact with the laser beam for a finite time before leaving the interaction region.
Since the number of atoms in a given volume element is statistically constant, this atom is replaced by a new atom from outside the beam.
The outgoing atom can be in any state, but the incoming atom is always in a ground state.
From the system's point of view, this represents an additional decay channel.
However, there are some notable differences from spontaneous emission.
Decay by spontaneous emission only occurs from excited states to dipole-allowed ground states with their individual rates $\Gamma_{ij}$.
In contrast, transit time broadening allows both excited and ground states to \enquote{decay}, with a uniform rate~\cite{sagle1996jpbamop}
\begin{equation}
\Gamma_\mathrm{t} = \frac{\bar{v}}{\bar{d}},
\end{equation}
where $\bar{d}$ is the average path length for an atom travelling through the beam and $\bar{v}$ the average velocity in the transverse plane, given by the Rayleigh distribution (the velocity probability distribution for particles in a plane).
We point out that this gives us the statistical average transit time rate. 
In general atoms have different velocities and cross the beam on different trajectories.
Nonetheless our simulations showed only marginal differences between taking the average decay rate and integrating over the different rates.

In the case of a circular beam, the transient decay rate is given by
\begin{equation}
    \Gamma_\mathrm{t} = \frac{\sqrt{8k_\mathrm B T/\pi m}}{D_\mathrm{FWHM}},
\end{equation}
where $D_\text{FWHM}$ is the diameter evaluated at half of the beam intensity~\cite{sagle1996jpbamop}, $m$ is the atomic mass in units of the kilogram, and $T$ is the temperature of the atomic vapour.
For a rubidium vapour at room temperature probed by a beam with diameter $D_\mathrm{FWHM}=\SI{1.0}{\mm}$, we obtain a decay rate of $\Gamma_\mathrm{t}\approx\SI{0.27}{\MHz}$, corresponding to an average interaction time of $\tau=\frac{1}{\Gamma_\mathrm{t}}\approx\SI{3.7}{\us}$.

The total decay rate $\Gamma_{ij}$ in \autoref{eq:lindblad_operator} between state $i$ and state $j$ is then the sum of spontaneous decay and transit time broadening:
\begin{equation}
    \Gamma_{\text{tot},ij} = \Gamma_{\text{nat},ij} + \frac{1}{2g_n}\Gamma_\mathrm{t}.
\end{equation}
This also implies an additional broadening of the spectral linewidth.
Note that we have introduced the factor $\frac{1}{2g_n}$, where $g_n$ denotes the degeneracy of the initial state.
This ensures that the coherence terms of the resulting Lindblad operator are correctly normalised.

The transverse particle motion leads to a transit-time broadening.
In contrast, the longitudinal velocity distribution $P(v_z)$ leads to a Doppler-broadened susceptibility $\chi_\mathrm{D}$, described by the convolution 
\begin{subequations}
    \begin{align}
        \chi_\text{D}(\Delta) &= \int\mathrm{d} v_z\,\chi(\Delta-kv_z)\,P(v_z)\\
         &= \int\mathrm{d} v_z^\prime\,\chi(kv_z^\prime)\,P(\frac{\Delta}{k}-v_z^\prime),
    \end{align}
\end{subequations}
where $k=2\pi/\lambda$ and $\Delta$ is the detuning from the atomic resonance frequency in the absence of the hyperfine splitting.
The substitution $v_z^\prime = \Delta/k-v_z$ reduces the computational burden, as $\chi(v_z^\prime)$ must only be computed once and can be reused for multiple values of $\Delta$.
\subsection{Propagation effects}
In contrast to the weak-probe theory, another effect  taken into account in this work is the attenuation of the laser light during its propagation through the rubidium vapour.
Here we used an iterative approach to calculate the absorption after passing through the \SI{2}{\mm} long vapour cell.
We divided the cell into 10 equal slices, along the laser propagation axis, such that we can consider ten vapour cells each of length \SI{0.2}{\mm} and a constant incident laser beam intensity along their length.
The intensity is then given by \cite{haupl2022njp}
\begin{equation}
    \frac{\mathrm{d}I}{\mathrm{d}z}=\frac{4\pi}{\lambda}\mathrm{Im}\left\{\sqrt{1+\chi\left[I(z), \Delta, n\right]}\right\}\,I(z),
    \label{eq:beer_lambert}
\end{equation}
where $\chi\left[I(z), \Delta, n \right]$ is the electric susceptibility.
Note that $\chi$ not only depends on the laser detuning $\Delta$ and the atomic density $n$, but also on the longitudinal beam intensity $I(z)$.
The resulting spectrum can be obtained by summing the transmission spectra of all slices.
We noticed that for 10 slices our simulation converges to an acceptable level of accuracy.
For a larger slice number, the computation would take longer while only producing a minor increase in accuracy.
Note that the required number of slices $N$ will vary depending on the absorption $\alpha$, cell length $L$ and atomic species (Na/K/Rb/Cs).
Currently, the value of $N$ required for the model to converge must be determined manually.
This can be easily done by manually setting $N$ as an argument in the main calculation.
Furthermore, the current model only allows the calculation of the propagation-corrected absorption for a single value of the detuning $\Delta$ at a time.
This is because for a given longitudinal position $z$ inside the medium, the intensity will change, depending on the detuning $\Delta$.
\subsection{Predicted spectrum}
\begin{figure}[ht]
    \centering
    \begin{tikzpicture}
    \node at (0,0) {\includegraphics[width=0.98\linewidth]{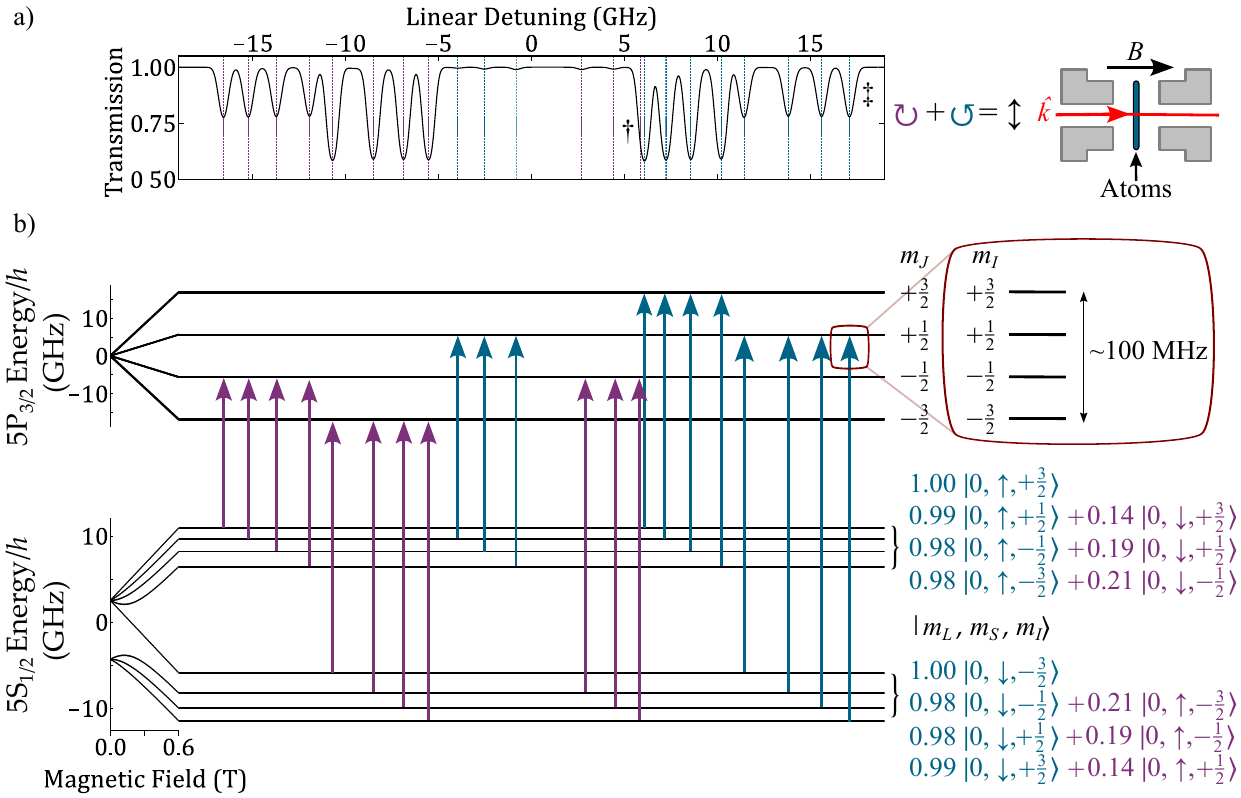}};
    \end{tikzpicture}
    \caption{The upper panel \textbf{(a)} shows a simulated spectrum of $^{87}$Rb  in the presence of a \SI{0.6}{\tesla} external magnetic field.
    The spectrum was computed using linearly polarised light, which is an equal sum of left and right circularly polarised light as indicated in the top right corner.
    The left of the lower panel \textbf{(b)} shows the energy level diagram of ground (5S$_{1/2}$) and excited (5P$_{3/2}$) states for a magnetic field increasing up to \SI{0.6}{\tesla}. The centre of the lower panel shows the energy levels at \SI{0.6}{\tesla}, and shares a frequency $x$-axis with panel \textbf{(a)}.  
    The purple ($\sigma^-$) and blue ($\sigma+$) arrows on \textbf{(b)} and vertical lines in \textbf{(a)} correspond to the transitions between the involved ground and excited states.
    The numbers in the bottom right corner give the involved states in the $\ket{m_L,m_S,m_I}$ basis, with small admixtures due to the hyperfine splitting at \SI{0.6}{\tesla}.
    The absorption lines in \textbf{(a)} marked with \dag{} and \ddag{} are taken as examples of closed and open transitions respectively, to investigate differences at high intensities.
    }
    \label{fig:big_diagram}
\end{figure}

A simulated spectrum of $^{87}$Rb for weak-probe intensities is shown in the upper panel of \autoref{fig:big_diagram}.
The lower panel shows the energy diagram of the ground and excited states, showing the evolution of the Zeeman shift for magnetic fields up to \SI{0.6}{\tesla}.
As expected from the discussion in Section~\ref{subsec:HPB}, \SI{0.6}{\tesla} is sufficiently large to gain access to the HPB regime, with the Zeeman splitting being larger than the hyperfine splitting.

The bottom right corner shows the involved ground states in the uncoupled basis $\ket{m_L,m_S,m_I}$. 
As expected for such a large field, $m_I$ is a good quantum number for the ground states; there are negligible admixtures due to the hyperfine interaction for the excited states. There are four $\sigma^+$ transitions (a consequence of the four different possible  values of $m_I$) from the  $m_S=+1/2$ manifold centred on a detuning of \SI{7.5}{\GHz}, and four $\sigma^+$ transitions  from the  $m_S=-1/2$ manifold centred on a detuning of \SI{15}{\GHz}. 
The separation of the transitions exceeds the Doppler width, yielding non-overlapping resonances. In addition there are three significantly weaker $\sigma^+$ transitions centred on a detuning of \SI{-2.5}{\GHz} arising from the small admixture of $m_S=-1/2$ in three of the upper four ground states. 
The pattern for the $\sigma^-$ transitions is mirror-symmetric about zero detuning.
Since the magnetic field is parallel to the light wave vector, $\sigma^+$/$\sigma^-$ transitions can be induced separately using a left/right-hand circularly polarised incident field (LCP/RCP)~\cite{steck2007}. 
In \autoref{fig:big_diagram}, the incident field is linearly polarised; this induces both $\sigma^+$/$\sigma^-$ transitions simultaneously since any linearly polarised light vector can be decomposed into an equal combination of LCP and RCP light~\cite{briscoe2024indirect}.

\section{Experimental setup}
\label{sec:experiment}
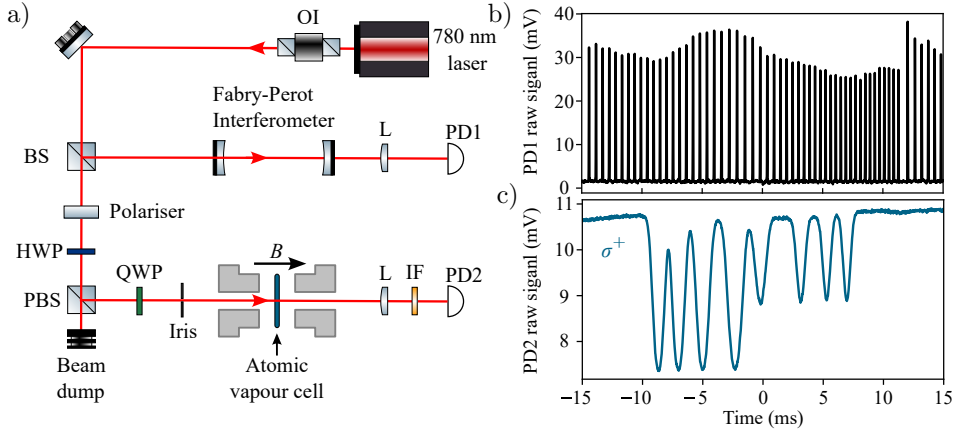
\begin{figure}[ht]
    \centering
    \begin{tikzpicture}
        \node at (0,0) {\includegraphics[width=0.98\linewidth]{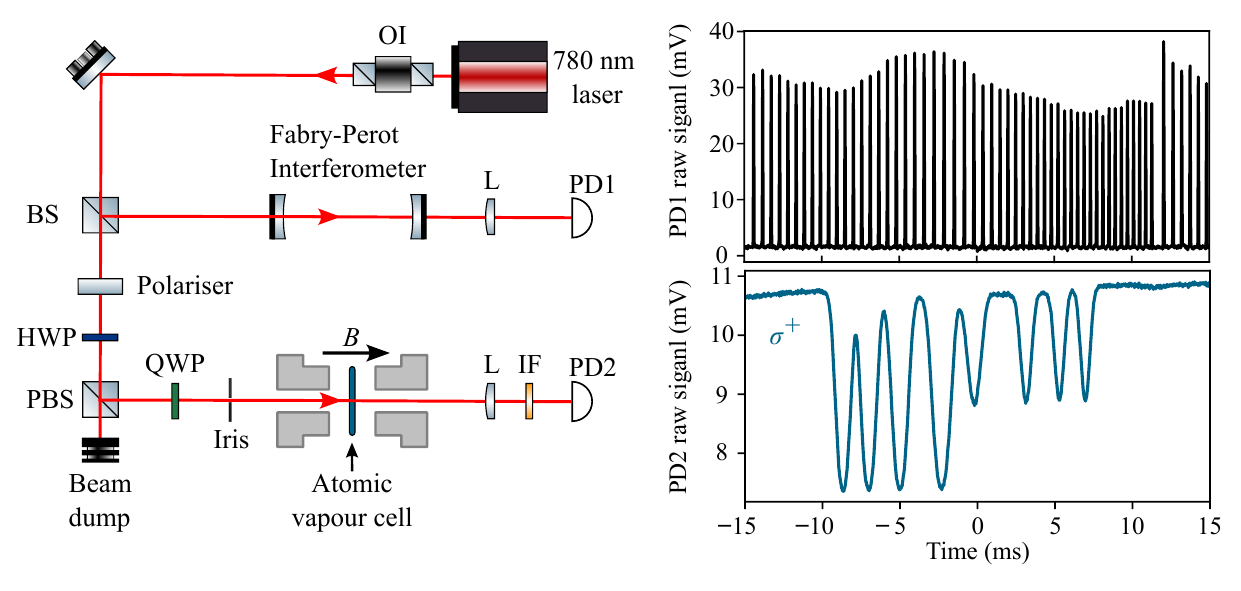}};
        \node at (-6.2,2.7) {a)};
        \node at (0.2,2.7) {b)};
        \node at (0.3,0.3) {c)};
    \end{tikzpicture}
    \caption{\textbf{(a)} Schematic depicting the experimental setup. After exiting the laser, the beam passes through an optical isolator OI before it is split via a beam splitter BS at a $90$:$10$ ratio. Some light goes through a Fabry-Perot interferometer to provide a frequency calibration~\cite{pizzey2022laser}. Most of the laser goes through a combination of polariser, half waveplate HWP and polarising beam splitter PBS to allow power control. A quarter waveplate QWP creates circularly polarised light.
    The beam diameter can be adjusted through an iris with diameters of respectively $0.5$, $1.0$ and \SI{2.0}{\mm}.
    A bandpass filter before the photodiode supresses stray light.
    The vapour cell is \SI{2}{\mm} long and surrounded by two top hat permanent magnets in Helmholtz geometry, resulting in a \SI{0.6}{\tesla} strong magnetic field at the centre of the cell.
    \textbf{(b)} shows an example of the recorded intensity of the photodiode PD1 as a function of time, when the laser frequency is scanned.
    \textbf{(c)} Example absorption spectrum recorded by the photodiode PD2.} 
    \label{fig:setup}
\end{figure}

To benchmark our model, we performed spectroscopy of the D2 line of $^{87}$Rb vapour under various conditions.
We utilised a thin vapour cell containing an isotope mixture of \SI{99.3}{\percent} $^{87}$Rb and \SI{0.7}{\percent} $^{85}$Rb. 
The cell thickness was \SI{2}{\mm}, to ensure a uniform magnetic field over the extent of the medium. 
Modest changes in temperature produce a large change in opacity of the cell~\cite{pizzey2022laser}, enabling experiments to be performed with either optically thick or thin media.
\autoref{fig:setup} shows a schematic of the setup. 
The beam of a Toptica DL100 diode laser operating at $\lambda=\SI{780}{\nm}$ is split into two. 
One passes through a Fabry-Perot interferometer to provide a frequency reference when scanning the laser frequency detuning. 
After a polariser, a combination of half-wave plate and polarising beam splitter controls the intensity of the second beam. 
Additional neutral density filters alter the beam power over almost 5 orders of magnitude, ranging from hundreds of microwatts up to tens of milliwatts.
The probe light is circularly polarised by means of a quarter-wave plate. 
With circular polarisation we obtain maximum absorption while ensuring that only one transition is addressed.
%

For this work we measured spectra for three different beam diameters that are controlled via an aperture in front of the vapour cell.
The original beam has a diameter of around $D_\mathrm{FWHM}=\SI{4.5}{\mm}$ before being restricted by interchangeable irises with diameters of $0.5$, $1.0$ and \SI{2.0}{\mm}. 
This gives us an approximately flat-top beam shape with precise diameters.
According to our simulations, the deviation from an ideal flat-top beam can be neglected in our case.
A flat intensity profile is advantageous, as it remains uniform along the cell during propagation.
In addition, we investigated the influence of the cell temperature.
We chose temperatures of $T\approx\SI{67}{\celsius}$ and $T\approx\SI{81}{\celsius}$, 
resulting in roughly 50\% and 80\% peak absorption respectively.
Note, the 80\% absorption attenuates the beam significantly.
Propagation effects can be neglected for $I < I_\mathrm{sat}$ or $I \gg I_\mathrm{sat}$, but need to be considered when the intensity is close to the saturation intensity.

The vapour cell is placed in an axial field of \SI{0.6}{\tesla} (oriented along the $z$-axis), produced by two cylindrical `top hat' NdFeB permanent magnets. The uniformity of the field across the length of the cell is better than 0.1$\%$~\cite{WhitingEIT}.

Revisiting \autoref{fig:big_diagram}, we note that the large  separation of states for a \SI{0.6}{\tesla} field allows us to separately address the \emph{closed} and \emph{open} transitions.
We  see that the \dag-marked transition is \emph{closed} as only one pure ground and one excited state are involved, effectively forming a two-level system.
In contrast, the open transition (\ddag) allows spontaneous decay into one or more \emph{dark states}. 
A dark state does not participate to a transition, but reduces the populations of the other states and thus the absorption.
For sufficient probe powers we expect the effect of optical pumping to affect these transition types differently.
\section{Results}
\label{sec:results}
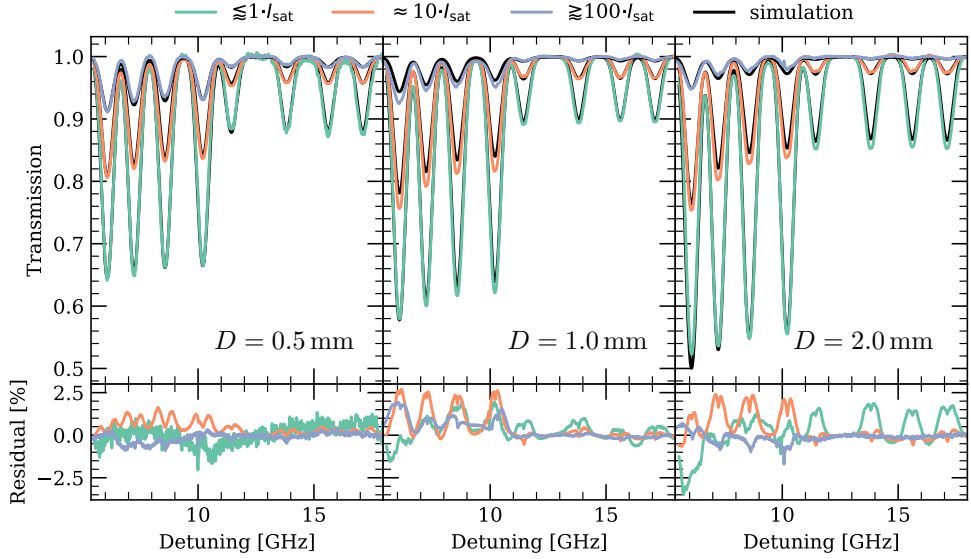
\begin{figure}[ht]
    \centering
    \begin{tikzpicture}
        \node at (0,0) {\includegraphics[trim=1cm 0 2cm 0, width=0.98\linewidth]{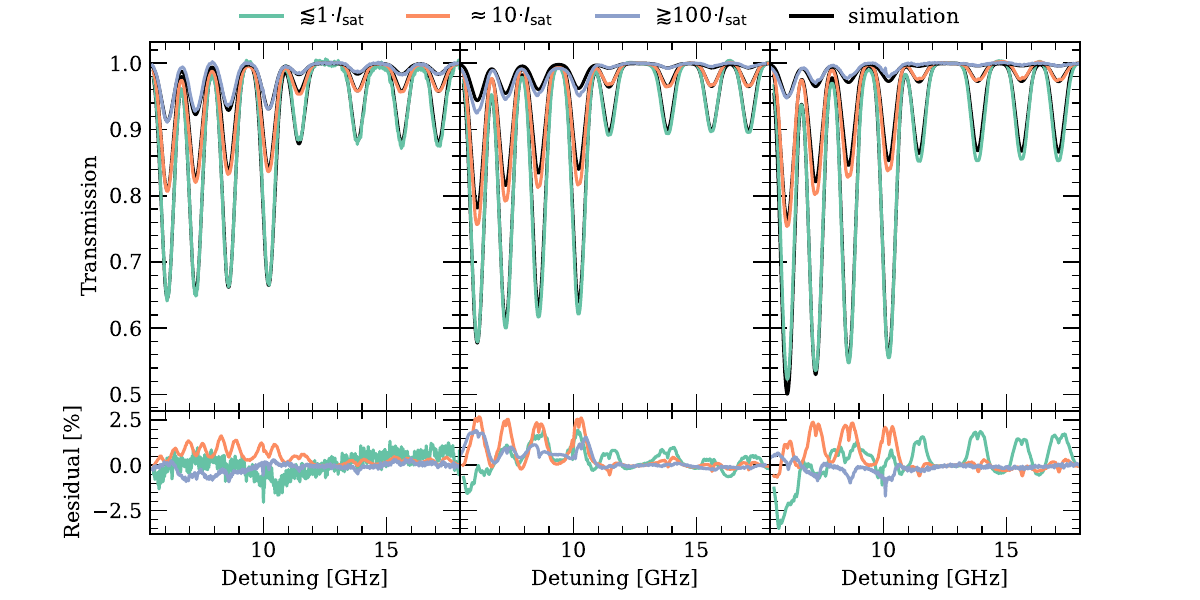}};
        \node at (-2.7,-0.8) {$D=0.5$\,mm};
        \node at (1.2,-0.8) {$D=1.0$\,mm};
        \node at (5,-0.8) {$D=2.0$\,mm};
    \end{tikzpicture}
    \caption{Measured spectra of $^{87}$Rb in the hyperfine Paschen-Back regime at \SI{0.6}{\tesla} for three different probe beam intensities and three different beam diameters.
    The intensities are chosen to be around $1\times$ (green lines), more than $10\times$ (orange lines) and more than $100\times$ the saturation intensity (purple lines).
    In comparison, the simulated spectra are shown in black, according to the experimental parameters.
    On the bottom we have plotted the residuals between experiment and simulation, showing very good agreement with less than 5\,\% difference over the entire optical range.
    }
    \label{fig:comparison_full_spectra}
\end{figure}

\begin{figure}[ht]
    \centering
    \includegraphics[width=\linewidth]{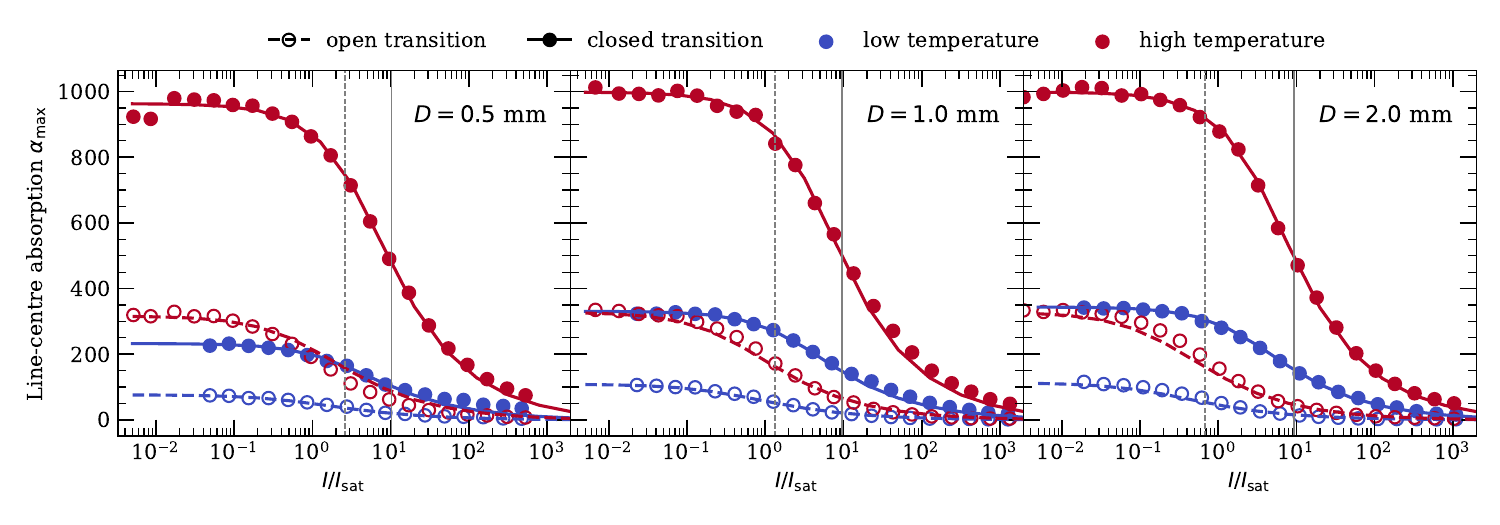}
    \caption{Evolution of the line-centre absorption $\alpha_\mathrm{max}$ of one closed (filled circles) and one open (hollow circles) transition as a function of the laser intensity.
    The transitions correspond to the those marked with \dag{} and \ddag{} in \autoref{fig:big_diagram}. Two different temperatures and the three different beam diameters are plotted.
    The vertical lines denote the laser intensities at which the absorption coefficients of the high temperature measurement are half of the values in the weak-probe limit.
    }
    \label{fig:comparison_center_linewidth_absorption}
\end{figure}

We measured the spectrum of the rubidium at 15 different intensity values. 
We have plotted exemplary spectra for three different intensity ranges in \autoref{fig:comparison_full_spectra}.
Not only can one see that the overall transmission increases with the intensity of the probe laser, but also that the relative line strengths change due to optical pumping.
Additionally, \autoref{fig:comparison_full_spectra} compares these measurements with our model (black lines).
The residuals plotted in the bottom row show very good agreement between our model and the experimental results.
Note that the intensity used for the respective simulated spectra were not obtained by fits to the individual measured spectra.
Instead, the intensity has been derived once by a global fit to the data, which is shown in \autoref{fig:comparison_center_linewidth_absorption}.
For higher intensities the residuals start showing dips on top of the absorption peaks.
We attribute this to back reflections from the uncoated cell windows, which produce saturation absorption spectroscopy (SAS) effects.
It has already been shown that SAS is possible for strong magnetic fields and circularly polarised light~\cite{sargsyan2015jetp}.

Another way to look at our data is by plotting the line-centre absorption coefficient $\alpha_\mathrm{max}$.
This is the maximum absorption coefficient for a given spectral line, related to the transmission $T$ via $\alpha=-\mathrm{ln}(T)/L$ with $L$ the cell length.
We selected the left-most peak (\dag, \autoref{fig:big_diagram}), which corresponds to a completely closed transition, and the right-most peak (open transition, \ddag, \autoref{fig:big_diagram}) for comparing the absorption as a function of the laser intensity.
The results are shown in \autoref{fig:comparison_center_linewidth_absorption}.
Here we have plotted $\alpha_\mathrm{max}$ of the open and closed transitions against the laser intensity, for two different temperatures and the three different beam diameters.
For most combinations of temperature and beam diameter we achieve good agreement between our experimental measurements and our model.
Note that most parameters for our model correspond to the experimental values, e.g. beam diameter, cell length or magnetic field strength. 
The model was fitted to the experimental data using only two free parameters, the cell temperature and attenuation of the laser intensity inside the cell. 
The attenuation corresponds to a shift of our model curve along the intensity axis.
However, the overall line shape is given by the fixed parameters.
The individual modelled spectra from \autoref{fig:comparison_full_spectra} are then directly produced using these parameters.

\autoref{fig:comparison_center_linewidth_absorption} also allows us to analyse the influence of transit time broadening.
When the beam diameter increases, so does the transit time, $\tau_t$, of atoms through the beam.
Therefore the decay rate is reduced ($\Gamma_\mathrm{t} = 1/\tau_t$).
For the high temperature data, this illustrated by the vertical lines, marking where the absorption coefficient is half the value of the weak-probe limit.
We see that the closed transition (filled dots) is only marginally affected by $\Gamma_\mathrm{t}$. 
Interestingly, the open transition (hollow dots) shows a clear shift towards higher intensities for smaller diameters (higher $\Gamma_\mathrm{t}$).

Both the data and model demonstrate that optical pumping causes saturation effects to set in at a lower intensity than would otherwise be expected~\cite{pappas1980saturation, smith2004role, sherlock2009weak}.
Furthermore, geometrical effects---especially the role of the probe beam diameter---must be taken into account when selecting a beam intensity to operate in the weak-probe regime, and
to obtain agreement between theory and experiment~\cite{sagle1996jpbamop, lindvall2007effect, lindvall2009interaction, moon2007theoretical, moon2008comparison}.
\section{Conclusion and Outlook}
\label{sec:outlook}
In summary, we have demonstrated an effective model for simulating the spectra of an alkali-metal vapour for a variety of experimental parameters, most distinctly at intensities in the saturation regime. 
While our model was only experimentally verified for the D2 line of rubidium, the software is also capable of simulating spectra of sodium, potassium and caesium over both D lines. 
In contrast to previous works~\cite{sagle1996jpbamop, bala2022jpbamop, bala2023competition}, we present line-centre absorption measurements over an intensity range of five orders of magnitude and for very high laser intensities of up to 1000 times the saturation intensity.
In addition, we showed individual atomic spectra for different intensities and compared them to our model.
Even at high magnetic field strengths reaching the hyperfine Paschen-Back regime, we can report very good agreement between experiment and simulations.
The good agreement can be maintained for very high laser intensity values, up to 1000 times the saturation intensity.
Furthermore, by using a strong magnetic field we enter the HPB regime which allows for a much easier comparison of the transit time broadening influence between open and closed  transitions. 
The model takes into account  changes in intensity due to propagation through the vapour and the effect of optical pumping.
Most importantly, the underlying software for the model is made publicly available at \cite{hauplg} to allow for its broad usage and integration into existing models. 
As the model is based on the optical Bloch equations, it is also possible to extract the state populations, e.g. for simulations based on fluorescence spectroscopy. 
In this investigation we used these populations to study the absorptive properties of an alkali-metal atomic medium, via the imaginary component of the electric susceptibility. 

In the future, we plan on extending the investigation to incorporate the dispersive properties of the medium, as encapsulated in the real component of the electric susceptibility. For applications such as designing narrowband optical filters at high fields, the interplay between the real and imaginary components is crucial~\cite{Faraday4, optimal, logue2022better}. 
Furthermore, the complex interaction mechanisms of buffer gas, like spin-exchange and velocity-changing collisions, warrant further theoretical modelling in the context of the optical Bloch equations.

\section{Acknowledgements}
This project is supported by the DFG (JO 1090/4-1, LO 1657/6-1 \& RU 1426/1-1) as part of the SPP 1929 \enquote{Giant Interactions in Rydberg Systems (GiRyd)}.
The authors also acknowledge EPSRC (Grant No. EP/R002061/1) for funding this work.
We also thank the IMPRS Physics of Light and SAOT for financial support. 

\section{Disclosures}
The authors declare no conflicts of interest.

\section{Data availability}
Data underlying the results presented in this paper are available from \cite{hauplg} and \cite{haupl2024}.

\newpage
\printbibliography
\end{document}